\documentclass[letterpaper, 10 pt, conference]{ieeeconf}  

\usepackage{ACSD-CR}
\usepackage{cite}
\usepackage{amsmath,amssymb,amsfonts}
\usepackage{algorithmic}
\usepackage{graphicx}
\usepackage{algorithm,algorithmic}
\usepackage{hyperref}
\hypersetup{hidelinks=true}
\usepackage{textcomp}

\IEEEoverridecommandlockouts                              

\usepackage{tikz}
\usetikzlibrary{calc}

\newcommand{\D}{\ensuremath{\mathcal D}}

\usepackage{xcolor}
\usepackage{xspace}
\usepackage{soul}
\usepackage{mathtools}
\usepackage{makecell}
\usepackage{cite}
\usepackage{nicefrac}
\usepackage{tcolorbox}
\usepackage{tabularx}
\usepackage{multirow}

\newcommand{\ra}{\rightarrow}

\newcommand{\ie}{\unskip, i.\,e.,\xspace}

\newcommand{\sut}{\text{s.\,t.\,}}

\newcommand{\N}{\ensuremath{\mathbb N }}

\newcommand{\R}{\ensuremath{\mathbb R }}

\newcommand{\X}{\ensuremath{\mathbb X }}

\newcommand{\U}{\ensuremath{\mathbb U }}

\newcommand{\co}{\ensuremath{\overline{\text{co}}}}

\newcommand{\eps}{\ensuremath{\varepsilon}}

\newcommand{\ball}{\ensuremath{\mathcal B}}

\definecolor{dgreen}{rgb}{0.0, 0.5, 0.0}

\newcommand{\spc}{\ensuremath{\,\,}}
\DeclareMathOperator*{\argmin}{arg\,min}

\DeclareMathOperator*{\arginf}{arg\,inf}

\newcommand{\norm}[1]{\left\lVert#1\right\rVert}  
\newcommand{\abs}[1]{\left\lvert#1\right\rvert}
\newcommand{\scal}[1]{\left\langle#1\right\rangle}
\newcommand{\set}[1]{\mathbb #1}

\makeatletter
\newcommand{\subalign}[1]{%
	\vcenter{%
		\Let@ \restore@math@cr \default@tag
		\baselineskip\fontdimen10 \scriptfont\tw@
		\advance\baselineskip\fontdimen12 \scriptfont\tw@
		\lineskip\thr@@\fontdimen8 \scriptfont\thr@@
		\lineskiplimit\lineskip
		\ialign{\hfil$\m@th\scriptstyle##$&$\m@th\scriptstyle{}##$\crcr
			#1\crcr
		}%
	}
}
\makeatother

\newtheorem{thm}{Theorem}
\newtheorem{dfn}{Definition}
\newtheorem{lem}{Lemma}

\newtheorem{asm}{Assumption}
\newtheorem{rem}{Remark}

\newcommand{\LL}{\text{Lip}_\text L}
\newcommand{\Lf}{\text{Lip}_\text f}

\title{\LARGE \bf
Some remarks on robustness of sample-and-hold stabilization
}

\author{
\thanks{\CRIEEE{10.1109/LCSYS.2025.3546990}{IEEE Control Systems Letters}{}}
Patrick Schmidt$^{1}$, Pavel Osinenko$^{2}$ and Stefan Streif$^{1}$
\thanks{$^{1}$P. Schmidt and S. Streif are with the Technische Universit\"at Chemnitz, Automatic Control and System Dynamics Lab, 09126 Chemnitz, Germany
        {\tt\small \{patrick.schmidt, stefan.streif@etit.tu-chemnitz.de\}}}%
\thanks{$^{2}$P. Osinenko is with Skolkovo Institute of Science and Technology, 143026 Moscow, Russia
        {\tt\small p.osinenko@yandex.ru}}%
}

\begin{document}

\maketitle
\thispagestyle{empty}
\pagestyle{empty}

\begin{abstract}
	This work studies robustness to system disturbance and measurement noise of some popular general practical stabilization techniques, namely, Dini aiming, optimization-based stabilization and inf-convolution stabilization.
	Common to all these techniques is the explicit usage of a (general nonsmooth) control Lyapunov function, thus allowing to see them as a kind of generalization to the celebrated Sontag's formula.
	It turns out that certain details of the above described robustness properties have not yet received the attention in literature they deserved.
	We provide new remarks, formalized in mathematical propositions, on robustness of selected popular stabilization techniques along with an extensive statistical case study on a robot parking problem.

\end{abstract}


\section{Introduction} \label{sec:intro}

There exists a great variety of stabilization algorithms for general nonlinear systems that explicitly use a control Lyapunov function (CLF) \cite{Rifford2002-semiconcave}, \cite{Sontag1999-clocks}. 
Such is the case in approaches that use the celebrated Sontag's formula \cite{Sontag1989-formula} with its numerous extensions, e.~g., to the case of adaptive control as highlighted in this seminal work \cite{Sepulchre2012-constructive}. 
Several prominent techniques such as Dini aiming (DiA), optimization-based control (OBC), and stabilization via infimum convolutions (InfC) give up the CLF smoothness requirement and render stabilization algorithms extending over the Sontag's formula.

These techniques along with optimization-based algorithms and steepest descent were investigated in e.g. \cite{Braun2017-SH-stabilization-Dini-aim}, \cite{Kellett2004-Dini-aim}, and \cite{Kellett2000-Dini-aim}.
Omitting the smoothness requirement on the one hand yields a great flexibility for their application, since as it is known, existence of a smooth CLF is rather a rare case in the first place (see \cite{Artstein1983-stabilization}, \cite{Brockett1983-stabilization}).
But on the other hand, the pay comes with the fact, that the thereby derived algorithms generally do not yield continuous feedback laws.
This has consequences for the robustness properties of the resulting controllers as highlighted in the literature.

The current work dives into the described matter of robustness via the following selected CLF-based stabilization algorithms: DiA, OBC, InfC. 
To give a breakdown, the contributions made here are the following: 
a study of robustness properties w.~r.~t. measurement errors (ME) and system disturbances (SD) of OBC is highlighted in Lemma \ref{lem:opt-bsd-ctrl-rob} along with extensive statistical simulation experiments on a mobile robot parking problem.
Also a new practical stabilization proof by OBC from the first principles is presented.
Although the statement of this result is not solely new, and indeed could be found in \cite{Braun2017-SH-stabilization-Dini-aim}, the latter work derived the proof based on the DiA technique. 
It should be noted that the robustness results differ for DiA, OBC, and InfC, which is also verified in our case study. 
Hence, we see Theorem \ref{thm:stabilization-OBC} as a particular contribution. 
Finally, for the reader's convenience, we present a complete exposition of the proof of DiA in Theorem \ref{thm:stabilization-Dini-aiming}, since the canonical result \cite{Kellett2002-advances} is unfortunately not easily accessible for the reader.
Unlike the work \cite{Kellett2002-advances}, our constructions are explicit without resorting to $\mathcal O$ terms.
Lemmas \ref{lem:conv-target-ball} and \ref{lem:InfC-rob} on practical stabilization and, respectively, robustness of the InfC algorithm are given forcompleteness.

The paper is organized as follows: 
Section \ref{sec:prob-descr} outlines the problem and introduces the necessary preliminaries.
Section \ref{sec:nonsmooth-stab-techniques} discusses the three stabilization methods in detail.
Their robustness w.r.t. MEs and SDs is investigated in Section \ref{sec:rob-nonsmooth-stab-techniques}.
Section \ref{sec:case-study} presents the case studies comparing the robustness of the three methods using the extended nonholonomic double integrator.
Finally, Section \ref{sec:concl} concludes the paper.

\textit{Notation}: $\ball_R(x)$ describes a ball with radius $R$ centered at the current state $x$ \ie $\ball_R(x) := \{ x: \norm x \leq R \}$, while $\ball_R$ means that $x = 0$; $\norm \cdot$ denotes the Euclidean norm; $\R_{\ge 0}$ refers to the set of non-negative real numbers; the state at time $t = k \delta$ is given as $x_k := x(k \delta)$, where $\delta$ is given as the sampling period and $k \in \N$; the closure of the convex hull of a set $\set A$ is given as $\co(\set A)$.

\section{Problem description} \label{sec:prob-descr}

In this work, uncertain nonlinear systems
\begin{equation}
	\label{eqn:sys}
	\dot x = f(x, u) + q, \quad y = \hat x := x + e,
\end{equation}
are considered, where $x \in \R^n$ is given as the state and $y$ is its measurement, which is used to determine the input $u = \kappa(\hat x)$. 
The (time-varying) disturbance is given as $q: \R_{\ge 0} \ra \R^n$.
It is assumed that the admissible control actions are in some compact input constraint set $\U$ (cf. \cite{Clarke1997-stabilization}) and that the following mild assumptions hold (cf. \cite{Schmidt2021-inf}):
\begin{asm}[System properties]
	\label{asm:sys-props}
	\hspace{1pt}
	\begin{itemize}
		\item (Disturbance boundedness) there exist numbers $\bar e, \bar q$ \sut $\forall t \ge 0$ $\norm{e(t)} \le \bar e$ and $\norm{q(t)} \le \bar q$;
		\item (Lipschitz property \cite{Clarke1997-stabilization}) for any $z \in \R^n$ and $\omega > 0$ there exists $\Lf > 0$ such that for all $x, y \in \ball_\omega(z)$ and for all $u \in \U: \norm{ f(x, u) - f(y, u) } \le \Lf \norm{x-y}$.
	\end{itemize}
\end{asm}
The three stabilization techniques rely on a proper, positive-definite, locally Lipschitz continuous control Lyapunov function (CLF) $L: \R^n \ra \R$ with Lipschitz constant $\LL$ obtained on a compact set $\set X$ and two functions $\alpha_1, \alpha_2 \in \mathcal K_\infty$ such that $\alpha_1(\norm x) \leq L(x) \leq \alpha_2(\norm x)$ holds for all $x \in \R^n$ \cite{Khalil2002-nonlin-sys}. 
There exist several techniques to determine (nonsmooth) CLFs, both numerical \cite{Baier2014-num-CLF}, \cite{Giesl2015-review} and analytical approaches \cite{Bianchini2018-merging}, \cite{Malisoff2009-constructing-strict-LF}, whereby we assume it as given.
For $L$, the \emph{lower directional generalized derivative} (LDGD) in a direction $\theta \in \R^n$ is defined as \cite{Clarke2000-feedback}
\begin{equation} 
	\label{eqn:dini-der}
	\D_\theta L(x) \triangleq \liminf_{ \mu \ra 0^+ } \frac{ L(x + \mu \theta) - L(x) }{\mu}.
\end{equation}
Let $w: \R^n \ra \R$ be a continuous non-negative function with $x \ne 0 \implies w(x) > 0$.
The LDGD satisfies \cite{Clarke1997-stabilization}: for each compact set $\X \subseteq \R^n$, there exists a compact $\U(\X) \subseteq \U$, s.t.
\begin{equation} 
	\label{eqn:decay}
	\forall x \in \X \inf_{ \theta \in \co (f(x, \U(\X))) } \D_\theta L(x) \le -w(x).
\end{equation}
Once the control law $\kappa$ is determined via the CLF, it is implemented in the sample-and-hold (SH) mode, which means the determined control is kept constant over the predefined sampling period $\delta$.
However, the approach is not limited to constant sampling periods (cf. \cite{Clarke1997-stabilization}).
The system description in the SH mode reads as
\begin{equation} 
	\label{eqn:sys-SH}
	\begin{split}
		\dot x = f(x, u_k) + q, t \in [ k \delta, (k + 1) \delta), u_k \equiv \kappa(\hat x (k \delta))
	\end{split}
\end{equation}
for $k \in \N$.
Applying a SH control, asymptotical stabilization is replaced by practical stabilization defined as \cite{Clarke2000-feedback}, \cite{Schmidt2021-inf}:
\begin{dfn}[Semiglobal robust practical SH stabilization] 
	\label{def:robust-pract-stab}
	A control law is said to robustly practically stabilize \eqref{eqn:sys} in the SH mode \eqref{eqn:sys-SH} if, for each $S$ and $s \in (0, S)$, there exist numbers $\tilde e = \tilde e(s, S) > 0$, $\tilde q = \tilde q (s, S) > 0$,  $\tilde \delta = \tilde \delta (s, S) > 0$
	depending uniformly on $s,S$ and $x \in \R^n$, such that if
	\begin{itemize}
		\item the sampling period satisfies $\delta \le \tilde \delta$ and
		\item the bounds on the ME and SD satisfy $\bar e \leq \tilde e$ and $\bar q \leq \tilde q$,
	\end{itemize}
	then, any closed-loop trajectory $x(t), \spc t \ge 0$, $x(0) = x_0 \in \ball_S$ is bounded and there exists $T$, s.t. $x(t) \in \ball_s, \spc \forall t \ge T$.
\end{dfn}

\begin{rem}
Practical SH stabilization can be considered as semi global as remarked in \cite{Clarke1997-stabilization} due to the fact that in particular $S$ can be chosen arbitrarily large, yet when it is fixed the bound on the sampling time is also fixed.
\end{rem}

Definition \ref{def:robust-pract-stab} introduces $S$ and $s$ as the radii of a starting ball and a target ball (see \cite[Fig. 3]{Schmidt2021-inf}).
The respective balls are defined based on some a priori bounds for $\bar e$ and $\bar q$, denoted as $\hat e$ and $\hat q$, e.g. $\hat e = \hat q = \frac{s}{8}$ (cf. \cite{Schmidt2021-inf}) ensuring $\hat s \geq 0$ and making them necessary to compute the minimum decay.
The later investigated stabilization methods follow a similar structure: first, a stabilizing control law is computed ensuring all trajectories starting in $\ball_S$ are uniformly bounded by an overshoot bound $S^\star$ defined as $S^\star := \alpha_1^{-1}(\hat L)$, where $\hat L := \sup_{\norm x \leq S} L(x)$.
The maximum CLF value within $\ball_{S^\star}$ is given as $\hat L^\star := \sup_{\norm{x} \leq S^\star} L(x)$.
Also the Lipschitz constant $\LL$ of $L$ is computed on $\ball_{S^\star}$, since the closed-loop trajectories are located in it.
The determined control law ensures a decay of the CLF between two sampling points until a core ball with radius $s^\star < s$ is reached obtaining a minimum decay for all $x \not\in \ball_{s^\star}$.
Its radius is defined as $s^\star := \alpha_2^{-1}(\frac{\hat \ell}{4})$, where $\hat \ell := \alpha_1(\hat s)$ and $\hat s = s - \hat e - \hat q$.
Second, it is shown that all trajectories enter a target ball $\ball_s$, $s > s^\star$ and remain in it, based on this decay.

The second part of this procedure is common to all methods and is used to truncate the following proofs (cf. in \cite[Theorem 1]{Schmidt2021-inf}).
\begin{lem} \label{lem:conv-target-ball}
	Consider system \eqref{eqn:sys-SH} with CLF $L$ along with a corresponding decay rate $w$ and the two functions $\alpha_1, \alpha_2 \in \mathcal K_\infty$.
	Let Assumption \ref{asm:sys-props} hold, $\delta$ be the sampling period, and $s$ be given as the radius of the target ball.
	Consider the definition of the different balls $\ball_{s^\star}, \ball_{S^\star}$ in Section \ref{sec:prob-descr} and let the minimum decay rate be given as 
	\begin{equation}
		\bar w := \inf_{\frac{s^\star}{2} \leq \norm x} w(x).
	\end{equation}
	If $L(x_{k + 1}) - L(x_k) \leq - \beta \delta \bar w$ holds for some $\beta > 0$ with $x_k := x(k \delta)$, then the trajectory $x(t; t_0, x_0)$ starting at $t = t_0$ in $x_0$ converges into $\ball_s$ and remains there for all $t \geq T$, where $T$ is given as the convergence time.
\end{lem}
The proof is given in the last part of the proof in \cite[Theorem 1]{Schmidt2021-inf}, where w.l.o.g. $\beta = \frac 3 8$ is chosen.

Since the preliminaries are completed, the stabilization techniques are first investigated in detail for 
\begin{equation}\label{eqn:sys-nominal}
	\dot x \in f(x,u)
\end{equation}
with $x(0) = x_0$ before \eqref{eqn:sys-nominal} is given up in favor of \eqref{eqn:sys} to study the robustness w.r.t. MEs and SDs.
These techniques can be also applied to differential inclusions, see, for example, \cite{Braun2017-SH-stabilization-Dini-aim}.

\section{Nonsmooth stabilization techniques} \label{sec:nonsmooth-stab-techniques}

\subsection{Dini Aiming} \label{sec:Dini-nominal}

The technique is presented in \cite{Kellett2000-Dini-aim} and follows a two-step procedure.
First, a local minimum of the CLF is computed on $\ball_r(x)$ for some $r > 0$, which is located at the boundary of $\ball_r(x)$ \cite[Lemma 5.2]{Kellett2002-advances}, i.e. $\partial\ball_r(x)$.
This leads to
\begin{equation} \label{eqn:Dini-aim-step-1}
	z^\star \in \argmin_{z \in \partial\ball_r(x)} L(z).
\end{equation}		
Second, the control law is determined based on the suitable decay direction $x - z^\star$ as
\begin{equation} \label{eqn:Dini-aim-step-2}
		\kappa(x) \in \argmin_{u \in \U} \scal{f(x,u), x - z^\star}.
\end{equation}

The following theorem (cf. \cite{Kellett2002-advances}) shows practical stabilization by the proposed steps of DiA.
It requires upper bounds for $\delta$ and $r$, that are presented first.
\begin{rem} \label{rem:bounds-DiA}
	In \cite[Section 4]{Kellett2004-Dini-aim}, the upper bounds
	\begin{equation}
		\begin{split}
			\bar \delta &:= \min \left\{ T_1, \frac{\ell_2 - \ell_1}{\LL M}, \frac{\eps_3}{M}, \frac{\sqrt{a_2^2 + 4 a_1 a_3} - a_2}{2 a_1} \right\} \\
			\bar r &:= \min \left\{ \frac{\eps_2}{\LL}, \eps_3, \eps_4, \frac{c}{\Lf \LL} \right\},
		\end{split}
	\end{equation}
	for $\delta$ and $r$ were derived, where $T_1$ is chosen to satisfy $\frac{c T_1}{16 \LL} \leq r - \sqrt{r^2 - \frac{r c T_1}{4 \LL}}$ and $M := \tilde M + \frac{c}{2 \LL}$ with $\tilde M$ as a bound of the closed-loop dynamics and $c$ as a minimal decay.
	The remaining parameters are defined as $a_1:= M^2 \LL$, $a_2 := M(M + r \Lf)$, $a_3 := \frac{c r}{4 \LL}$ and the terms $\ell_1$, $\ell_2$, $\eps_2$, $\eps_3$, $\eps_4$ describe the set, on which the Lipschitz constants $\LL$ and $\Lf$ as well as the decay condition are computed, i.e. $\ball_{S^\star}$ in our case, where $S^\star$ is defined based on $S$.
\end{rem}
\begin{thm} \label{thm:stabilization-Dini-aiming}
	Consider system \eqref{eqn:sys-nominal} with CLF $L$, a corresponding decay rate $w$, the two functions $\alpha_1, \alpha_2 \in \mathcal K_\infty$ as well as $\bar \delta$ and $\bar r$ in Remark \ref{rem:bounds-DiA}.
	Let Assumption \ref{asm:sys-props} hold and let the control be defined as \eqref{eqn:Dini-aim-step-2} with $0 < r \leq \bar r$.
	Applying $\kappa(x)$ in the SH manner with sampling period $\delta \leq \bar \delta$ practically stabilizes the system at its origin for all $x(0) = x_0 \in \ball_S$.
\end{thm}
\begin{proof}
	The proof is divided into three parts.
	The first part connects the LDGD \eqref{eqn:dini-der} with the scalar product of the decay direction and $x_k - z^\star_k$.
	The second part determines the actual decay.
	The third part computes the reaching time to the target ball and shows that the trajectory remains in it.
	
	\textbf{Step 1}: Connection between LDGD and scalar product: \\
Let $\kappa(x_k)$ be the minimizer at time step $k$ according to \eqref{eqn:Dini-aim-step-2} and let the stabilizing direction be given as $\theta_k^\star = f(x_k, \kappa(x_k))$.
	Let $\eps > 0$ be small.
	Consider the line segment connecting $x_k$ and $z^\star_k$ (see Fig. \ref{fig:proof-pt-1}).
	Shift this line such that it passes through $z^\star_k + \eps \theta^\star_k$ while remaining parallel to the original line.
	Let $\hat z_k$ be the intersection point between the shifted line and $\partial \ball_r(x_k)$, which is closest to $z_k^\star$.
	By construction, it follows that $\hat z_k - (z_k^\star + \eps \theta_k^\star)$ and $x_k - z_k^\star$ are parallel and point in the same direction.
	Their scalar product is given as
	\begin{equation} \label{eqn:thm-proof-scalar-product-1}
		\begin{split}
			&\scal{\hat z_k - (z_k^\star + \eps \theta_k^\star), \frac{x_k - z_k^\star}{r}}
			= \norm{\hat z_k - (z_k^\star + \eps \theta_k^\star)} \\
			&\geq \frac{L(\hat z_k) - L(z_k^\star + \eps \theta_k^\star)}{\LL} \geq \frac{L(z_k^\star) - L(z_k^\star + \eps \theta_k^\star)}{\LL}
		\end{split}
	\end{equation}
	with $\norm{x_k - z_k^\star} = r$ since $z_k^\star \in \partial\ball_r(x_k)$.
	The last inequalities hold since $L$ is locally Lipschitz continuous with $\LL$ and since $L(z_k^\star) \leq L(z_k)$ holds for all $z_k \in \ball_r(x_k)$.
		\begin{figure}
	\centering
		\begin{tikzpicture}
			\begin{scope}[rotate = -30]
				\draw[scale=1, domain=-2:2, smooth, variable=\x, black!50!white] plot ({\x}, {-0.1*\x*\x-2}) node[above] {$L$};
				\draw[dotted] (0,0) circle (2cm);
				\draw[fill = black] (0,0) circle (0.02cm) node[right] {$x_k$};
				\draw[<-] (0,0) -- (0,-2) node[right, midway] {$x_k - z_k^\star$};
				\draw[fill = black] (0,-2) circle (0.02cm) node[right] {$z_k^\star$};
				\draw[fill = black] (-105:2) circle (0.02cm) node[anchor = south east] {$\hat z_k$};
				\draw[dashed] (-105:2) -- (105:2);
				\draw[->] (0,-2) -- (-105:2);
				\draw[->] (0,-2) -- ($(-105:2)+(0,-0.5)$);
				\draw[fill = black] ($(-105:2)+(0,-0.5)$) circle (0.02cm) node[left] {$z_k^\star + \eps\theta_k^\star$};
				\draw[->] ($(-105:2)+(0,-0.5)$) -- (-105:2);				
				\draw[dotted] (45:2) node[anchor = south west] {$\ball_r(x_k)$};
			\end{scope}
		\end{tikzpicture}
		\caption{Setting of the proof in step 1}
		\label{fig:proof-pt-1}
	\end{figure}
	Rearranging the terms in \eqref{eqn:thm-proof-scalar-product-1} yields
	\begin{equation}
		\begin{split}
			&\scal{\hat z_k - z_k^\star, \frac{x_k - z_k^\star}{r}} - \eps \scal{\theta_k^\star, \frac{x_k - z_k^\star}{r}} \\
			& \ \ \ \ \geq \frac{L(z_k^\star) - L(z_k^\star + \eps \theta_k^\star)}{\LL} \\
		\end{split}
	\end{equation}
	or, equivalently
	\begin{equation} \label{eqn:thm-proof-scalar-product-5}
		\begin{split}	
			&\scal{\theta_k^\star, \frac{x_k - z_k^\star}{r}} \\
			&\leq \underbrace{\frac{L(z_k^\star + \eps \theta_k^\star) - L(z_k^\star)}{\eps \LL}}_{=: A(\eps)} + \underbrace{\scal{\frac{\hat z_k - z_k^\star}{\eps}, \frac{x_k - z_k^\star}{r}}}_{=: B(\eps)}. \\
		\end{split}
	\end{equation}
	First, $B(\eps)$ in \eqref{eqn:thm-proof-scalar-product-5} is replaced by the scalar product according to $\scal{X,Y} \leq \norm X \norm Y \cos(\angle (X,Y))$ and the law of cosines for $\cos(\angle (X,Y))$ resulting in
	\begin{equation} \label{eqn:thm-proof-scalar-product-7}
		\begin{split}
			& B(\eps) \\
			&\leq \frac{\norm{\hat z_k - z_k^\star}}{\eps} \frac{\norm{\hat z_k - z_k^\star}^2 + \norm{x_k - z_k^\star}^2 - \norm{x_k - \hat z_k}^2}{2 \norm{\hat z_k - z_k^\star} \norm{x_k - z_k^\star}} \\
			&= \frac{\norm{\hat z_k - z_k^\star}}{\eps} \frac{\norm{\hat z_k - z_k^\star} + r^2 - r^2}{2 r} = \frac{\norm{\hat z_k - z_k^\star}^2}{2 r \eps}. \\	
		\end{split}
	\end{equation}
	Based on the construction of $\hat z_k$ (see Fig. \ref{fig:proof-pt-1}), all points on the line passing through it are given by $\{ z_k^\star + \eps \theta_k^\star + \lambda (x_k - z_k^\star), \lambda \in \R \}$.
	To determine the intersection point $\hat z_k$ of this line and $\ball_r(x_k)$, $\hat \lambda$ is obtained by solving:
	\begin{equation}
		\begin{split}
			&\norm{z_k^\star + \eps \theta_k^\star + \hat \lambda (x_k - z_k^\star) - x_k}^2 \stackrel{!}{=} r^2 \\
			&\Leftrightarrow \hat \lambda^2 \scal{x_k - z_k^\star, x_k - z_k^\star} + 2 \hat \lambda \scal{x_k - z_k^\star, z_k^\star + \eps \theta_k^\star - x_k} \\
			&\ \ \ \; + \scal{z_k^\star + \eps \theta_k^\star - x_k, z_k^\star + \eps \theta_k^\star - x_k} - r^2 \stackrel{!}{=} 0.
		\end{split}
	\end{equation}		
	Using the quadratic formula, two solutions $\lambda_1$ and $\lambda_2$ depending linearly on $\eps$ are obtained.
	Replacing $\hat z_k  = z_k^\star + \eps \theta_k^\star + \lambda_1 (x_k - z_k^\star)$ in \eqref{eqn:thm-proof-scalar-product-7} and applying $\liminf_{\eps \searrow 0}$ yields
	\begin{equation}
		\liminf_{\eps \searrow 0} B(\eps) \leq \liminf_{\eps \searrow 0} \frac{\norm{\eps \theta_k^\star + \lambda_1 (x_k - z_k^\star)}^2}{2 r \eps} = 0.
	\end{equation}
	Thus, applying $\liminf_{\eps \searrow 0}$ on \eqref{eqn:thm-proof-scalar-product-5} establishes the desired connection between the LDGD and the scalar product as
	\begin{equation} \label{eqn:thm-proof-scalar-product-6}
		\begin{split}
			\scal{\theta_k^\star, \frac{x_k - z_k^\star}{r}} &\leq \liminf_{\eps \searrow 0} A(\eps) + B(\eps) = \frac{1}{\LL} \D_{\theta_k^\star} L(z_k^\star).
		\end{split}
	\end{equation}
	
	\textbf{Step 2}: Determine decay: \\
	In this step, it remains to show that $L(x_{k + 1}) < L(x_k)$ holds.
	Let $\delta \leq \bar \delta$ hold.
	Then, $\delta$ is small enough such that the trajectory of the state moves almost along $\theta_k^\star = f(x_k, \kappa(x_k))$, which holds due to 
	\begin{equation}
		x_{k+1} = x_k + \int_{k\delta}^{(k + 1)\delta} f(x(\tau), \kappa(x_k)) \ \mathrm d \tau
	\end{equation}		
	with sampling period $\delta$ (see Fig. \ref{fig:proof-pt-2}).
	The construction of $\theta_k^\star$ ensures that it is a direction which guarantees a decrease w.r.t. the given CLF.	 
	Moving along $\theta_k^\star$ ensures $z_k^\star \in \ball_r(x_{k + 1})$, but the minimizer in \eqref{eqn:Dini-aim-step-1} is given as $z_{k + 1}^\star$.
	Therefore,
	\begin{equation}
		L(z_{k + 1}^\star) \leq L(z_k^\star + \eps \theta_k^\star) \leq L(z_k^\star)
	\end{equation}		
	 follows.
	Additionally, $\norm{x_{k + 1} - z_k^\star} \leq r = \norm{x_k - z_k^\star}$ holds.
	Since there exists a minimal decay at $z^\star$ outside the core ball, say $\bar w > 0$, there exists also a stabilizing $\theta$ such that $D_{\theta} L(z^\star) \leq - \bar w$ holds.
	Along with some $\eps_1 = \eps_1(x, \theta) > 0$, 	
	\begin{equation} \label{eqn:decay-theta-is-nu-half}
		\frac{L(z^\star + \eps_1 \theta) - L(z^\star)}{\eps_1} \leq - \frac{\bar w}{2}
	\end{equation}
	holds.	
	With \eqref{eqn:decay-theta-is-nu-half}, it follows that $L(z_k^\star + \eps \theta_k^\star) - L(z_k^\star) \leq - \eps \frac{\bar w}{2}$ holds and with $L(z_{k + 1}^\star) \leq L(z_k^\star + \eps \theta_k^\star)$, it follows that also $L(z_{k + 1}^\star) - L(z_k^\star) \leq - \delta \frac{\bar w}{2}$ holds for the sampling period $\delta$.
	A decay from $x_k$ to $x_{k+1}$ follows with $\abs{L(z_k^\star) - L(x_k)} \leq \LL \norm{z_k^\star - x_k}$ as
	\begin{equation}
		\begin{split}
			&L(x_{k + 1}) - L(x_k) \\
			&\leq \LL r + L(z_{k+1}^\star) - L(z_k^\star) + \LL r \\
			&\leq 2 \LL r - \delta \frac{\bar w}{2}.
		\end{split}
	\end{equation}
Choosing $r \leq \frac{\delta \bar w}{4 \LL}$ yields the desired decay as $L(x_{k + 1}) - L(x_k) \leq - \delta \frac{\bar w}{4}$.
	\begin{figure}
	\centering
		\begin{tikzpicture}
			\begin{scope}[rotate = -30]
				\draw[scale=1, domain=-2:2, smooth, variable=\x, black!50!white] plot ({\x}, {-0.1*\x*\x-2}) node[above] {$L$};
				\draw[scale=1, domain=-2:2, smooth, variable=\x, black!50!white] plot ({\x-1}, {-0.1*\x*\x-2.625}) node[above] {$L$};
				\draw[dotted] (0,0) circle (2cm);
				\draw[fill = black] (0,0) circle (0.02cm) node[right] {$x_k$};
				\draw[<-] (0,0) -- (0,-2) node[right, midway] {$x_k - z_k^\star$};
				\draw[fill = black] (0,-2) circle (0.02cm) node[right] {$z_k^\star$};
				\draw[white] (-105:2) -- (105:2) node[left, midway] {$\ball_r(x_k)$};
				\draw[->] (0,-2) -- ($(-105:2)+(0,-0.5)$) node[below, midway] {$\theta_k^\star$};
							
				\draw[->] (0,0) -- ($(-105:2)+(0,1.5)$) node[above, midway] {$\theta_k^\star$};
				\draw[dotted] (0,0) to[bend right = 10] ($(-105:2)+(0,1.3)$);
				\draw[fill = black] ($(-105:2)+(0,1.3)$) circle (0.02cm) node[left] {$x_{k+1}$};
				\draw[dotted] ($(-105:2)+(0,1.3)$) circle (2cm);
				\draw[fill = black] ($(-105:2)+(0,1.3)+(-95:2)$) circle (0.02cm) node[left] {$z_{k+1}^\star$};
				\draw[dotted] (45:2) node[anchor = south west] {$\ball_r(x_k)$};
				\draw[dotted] ($(-105:2)+(0,1.3)+(135:2)$) node[anchor = south east] {$\ball_r(x_{k + 1})$};
			\end{scope}
		\end{tikzpicture}
		\caption{Setting of the proof in step 2}
		\label{fig:proof-pt-2}
	\end{figure}
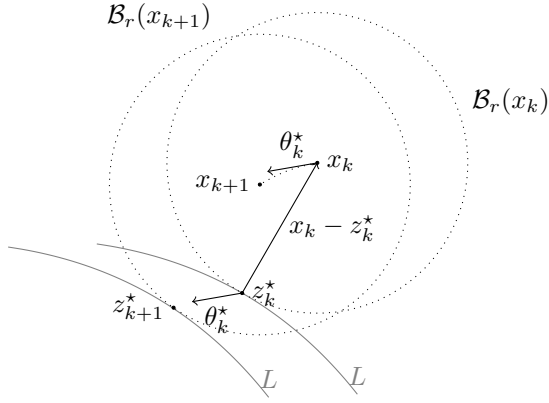
			
	\textbf{Step 3}: Reaching time and stay in target ball: \\
	Due to the determined decay of $- \delta \nicefrac{\bar w}{4}$ for $\norm{x} \geq s^\star$, Lemma \ref{lem:conv-target-ball} ensures that the trajectory remains in the desired region $\ball_s$. 
	Since $- \delta \frac{\bar w}{4}$ is obtained as a decay until the closed-loop trajectories reach $\ball_{s^\star}$, the reaching time is computed as $T := 4 \frac{\hat L^\star - \nicefrac{\hat \ell}{2}}{\delta \bar w}$ with $\nicefrac{\hat \ell}{2}$ as the minimum value and $\hat L^\star$ as the maximum value of the CLF based on $S^\star$ (cf. Section \ref{sec:prob-descr}).
\end{proof}

The algorithm is summarized in Algorithm \ref{alg:dini-aiming}.
\begin{algorithm}[H]
	\caption{Dini aiming}
	\label{alg:dini-aiming}
	\begin{algorithmic}[1]
		\renewcommand{\algorithmicrequire}{\textbf{Input:}}
		\renewcommand{\algorithmicensure}{\textbf{Set:}}
		\REQUIRE System $\dot x = f(x, u)$ and a CLF $L(x)$
		\ENSURE  Sampling period $\delta$
		\\ At $t_k = \delta k$:
			\STATE Compute $z^\star_k$ as \eqref{eqn:Dini-aim-step-1}
			\STATE Compute $\kappa$ as \eqref{eqn:Dini-aim-step-2}
			\STATE Apply $\kappa$ to the system and hold it constant until the next sample $k+1$
	\end{algorithmic}
\end{algorithm}

\subsection{Optimization-based control} \label{sec:opt-bsd-ctrl}

OBC combines the two steps \eqref{eqn:Dini-aim-step-1} and \eqref{eqn:Dini-aim-step-2} of DiA (cf. optimization-based feedback in \cite{Braun2017-SH-stabilization-Dini-aim}) as
\begin{equation} \label{eqn:opt-bsd-ctrl-law}
	\kappa(x) \in \argmin_{u \in \U} L(x + \delta f(x, u)).
\end{equation}
In contrast to \cite{Braun2017-SH-stabilization-Dini-aim}, where the closed-loop trajectory is integrated, here we use explicitly the forward Euler method to approximate the next value of the CLF provided by $\kappa(x)$.

For the stabilization result, we pose the following mild assumption, that was also used in \cite{Schmidt2021-inf}.
\begin{asm} \label{asm:homog-asm}
	The CLF $L$ is \emph{locally regular} and obtains a locally uniform LDGD \eqref{eqn:dini-der} meaning that for all compact sets $\set Y, \Theta \subset \R^n$ and for all $\nu, \chi > 0$, there exists $\tilde{\set Y} \subseteq \set Y$ and some $\mu \geq 0$, such that
	\begin{enumerate}
		\item for all $\tilde y \in \tilde{\set Y}, \theta \in \Theta, \mu' \in (0, \mu]$, it holds that 
		\begin{equation} \label{eqn:homog-asm-eq-1}
			\abs{\frac{L(\tilde y + \mu' \theta) - L(\tilde y)}{\mu'} - \D_\theta L(\tilde y)} \leq \nu.
		\end{equation}	
	\item for each $y \in \set Y$ there exists $\tilde y \in \tilde{\set Y}$ such that
	\begin{equation} \label{eqn:homog-asm-eq-2}
		\norm{ y - \tilde y } \leq \chi.
	\end{equation}
\end{enumerate}		
\end{asm}

Assumption \ref{asm:homog-asm} holds,~e.~g., for any function satisfying that for each compact $\X$, there exist $L_1, \ldots, L_p$ with $p < \infty$ and $\X_i$ such that $\text{supp}(L_i) := \overline{ \{ x \in \X: L_i(x) \not= 0 \}} = \X_i$ and $L = \sum_{i = 1}^p L_i$ on $\X$.
Intuitively, these are functions that are locally piece-wise smooth such as those resulting from triangulation-based numerical constructions of CLFs \cite{Baier2014-num-CLF}.
For some rather exotic function that one would not expect as a CLF, like $f_1(x) = \sin \left( \frac 1 x \right)$, the assumption does not hold due to an infinite number of oscillations as $x$ approaches zero.
In the following, it is shown that a decay is obtained by using \eqref{eqn:opt-bsd-ctrl-law}.
\begin{thm}[Practical stabilization of OBC] \label{thm:stabilization-OBC}
	Consider the setting of Theorem \ref{thm:stabilization-Dini-aiming} and let Assumption \ref{asm:homog-asm} hold.
	Applying \eqref{eqn:opt-bsd-ctrl-law} on \eqref{eqn:sys-nominal} in the SH mode practically stabilizes the system at its origin according to Definition \ref{def:robust-pract-stab}.
\end{thm}

\begin{proof}
	First, determine the decay from $t = k \delta$ to $t = (k + 1) \delta$.
	Let $\bar w > 0$ be given.
	There exists some direction $\theta_k^\#$ such that a decay of $-\bar w$ is obtained, i.e. $\D_{\theta_k^\#} L(x_k) \leq -\bar w$.
	Since \eqref{eqn:homog-asm-eq-1} in Assumption \ref{asm:homog-asm} holds,
	\begin{equation} \label{eqn:loc-uniform-dini-der}
		\frac{L(x_k + \delta \theta_k) - L(x_k)}{\delta} \leq \D_{\theta_k} L(x_k) + \frac{\bar w}{2}
	\end{equation}
	is locally fulfilled and since $\kappa(x_k)$ is given as \eqref{eqn:opt-bsd-ctrl-law}, 
	\begin{equation}
		L(x_k + \delta f(x_k, \kappa(x_k))) \leq L(x_k + \delta \theta_k^\#)
	\end{equation}		
	is satisfied.
	The desired result follows as 
	\begin{equation}
		\begin{split}
			&L(x_k + \delta f(x_k, \kappa(x_k))) - L(x_k) \\
			&\leq \delta \left( \D_{\theta_k^\#} L(x_k) + \frac{\bar w}{2} \right) \leq - \delta \frac{\bar w}{2}.	
		\end{split}
\end{equation}		
	On the other hand, if \eqref{eqn:homog-asm-eq-1} does not hold at the current $x_k$, there exists some point $\tilde x_k$ at which \eqref{eqn:homog-asm-eq-1} is satisfied and $\norm{x_k - \tilde x_k} \leq \chi$ holds. 
	Thus, 
	\begin{equation} \label{eqn:decay-with-homogenity-help}
		\begin{split}
			&L(x_k + \delta \theta_k) - L(x_k) \\
			&\leq 2 \LL \norm{\tilde x_k - x_k} + L(\tilde x_k + \delta \theta_k) - L(\tilde x_k) \\
			&\leq 2 \LL \chi + \delta \left( \D_{\theta_k^\#} L(x_k) + \frac{\bar w}{2} \right) \leq 2 \LL \chi - \delta \frac{\bar w}{2}. 
		\end{split}
	\end{equation}
	If $\chi$ is chosen as $\chi \leq \frac{\delta \bar w}{8 \LL}$, \eqref{eqn:decay-with-homogenity-help} reads as 
	\begin{equation}
		L(x_k + \delta \theta_k) - L(x_k) \leq 2 \LL \chi - \delta \frac{\bar w}{2} \leq - \delta \frac{\bar w}{4}.
	\end{equation}
	This shows that even if \eqref{eqn:homog-asm-eq-1} in Assumption \ref{asm:homog-asm} does not hold at the current $x_k$, there exists still a decay due to \eqref{eqn:homog-asm-eq-2}.	
		The convergence of the closed-loop trajectory into the target ball is ensured by Lemma \ref{lem:conv-target-ball}. 
\end{proof}

\subsection{Infimum-convolution stabilization} \label{sec:inf-con-stab}

A third stabilization technique is based on the so-called infimum convolution of the CLF $L$ given as 
\begin{equation}
	L_\alpha(x) = \inf_{y \in \R^n} \left\{ L(y) + \frac{1}{2 \alpha^2} \norm{y - x}^2 \right\}
\end{equation}
for some $\alpha \in (0,1)$.
The smaller $\alpha$ is, the closer the approximation $L_\alpha$ is to $L$.
A minimizer of $V_\alpha(x)$ is denoted as
\begin{equation} \label{eqn:InfC-minimizer}
	y_\alpha(x) \in \arginf_{y \in \R^n} \left\{ L(y) + \frac{1}{2 \alpha^2} \norm{y - x}^2 \right\}.
\end{equation}
It is used to define a proximal subgradient as (cf. \cite{Clarke1997-stabilization}).
\begin{equation} \label{eqn:prox-subgr}
	\zeta_\alpha(x) = \frac{x - y_\alpha(x)}{\alpha^2}.
\end{equation}
The resulting control law is computed by minimizing the decay at the current point $x$ using $\zeta_\alpha(x)$ as a descent direction:
\begin{equation} \label{eqn:InfC-ctrl-law}
	\kappa(x) \in \arginf_{u \in \U} \scal{\zeta_\alpha(x), f(y_\alpha(x),u)}.
\end{equation}
In \cite{Schmidt2021-inf} it is shown that \eqref{eqn:InfC-ctrl-law} practically stabilizes \eqref{eqn:sys} even in the presence of MEs and SDs. 

\section{Robustness of nonsmooth stabilization techniques} \label{sec:rob-nonsmooth-stab-techniques}

From now on, system \eqref{eqn:sys} is considered and bounds for the ME and SD are derived to ensure practical stability.

\subsection{Robustness of Dini Aiming} \label{sec:rob-Dini-aiming}

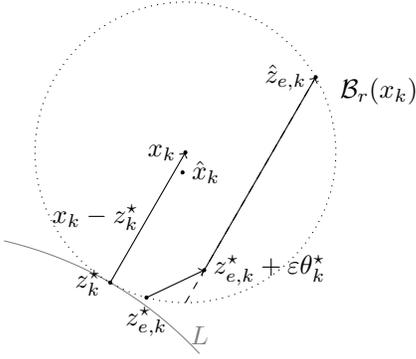
\begin{figure}
	\centering
		\begin{tikzpicture}
			\begin{scope}[rotate = -30]
				\draw[scale=1, domain=-1.5:1.5, smooth, variable=\x, black!50!white] plot ({\x}, {-0.1*\x*\x-2}) node[above] {$L$};
				\draw[dotted] (0,0) circle (2cm);
				\draw[fill = black] (0,0) circle (0.02cm) node[left] {$x_k$};
				\draw[fill = black] (0.1,-0.25) circle (0.02cm) node[right] {$\hat x_k$};
				\draw[<-] (0,0) -- (0,-2) node[left, midway] {$x_k - z_k^\star$};
				\draw[fill = black] (0,-2) circle (0.02cm) node[left] {$z_k^\star$};
				\draw[fill = black] (-75:2) circle (0.02cm) node[below] {$z_{e, k}^\star$};
				\draw[fill = black] (60:2) circle (0.02cm) node[left] {$\hat z_{e, k}$};
				\draw[dashed] (-60:2) -- (60:2);
				\draw[->] ($(-60:2) + (0,0.5)$) -- (60:2);
				\draw[fill = black] ($(-60:2) + (0,0.5)$) circle (0.02cm) node[right] {$z_{e, k}^\star + \eps\theta_k^\star$};
				\draw[->] (-75:2) -- ($(-60:2) + (0,0.5)$);				
				\draw[dotted] (45:2) node[anchor = south west] {$\ball_r(x_k)$};
			\end{scope}
		\end{tikzpicture}
		\caption{The difficulty of Dini aiming along with measurement errors.}
		\label{fig:proof-opt-inac}
	\end{figure}

It is difficult deriving a bound on the ME in DiA that does not depend on the sampling period and thus on the stabilization method.
The reason for this lies in the computation of the minimizer in \eqref{eqn:Dini-aim-step-1} for a measurement $\hat x_k$.
Using this minimizer to compute a decay direction $\hat \theta_k^\star$ and applying it to $x_k$ could yield an inward-pointing vector, such that the value of the CLF increases (see Fig. \ref{fig:proof-opt-inac}).
Such an issue is generally difficult to resolve.

In \cite[Section 4.4]{Kellett2004-weak}, the authors investigated robustness of Dini aiming w.r.t. a SD, i.e. $\dot x = f(x, u) + d$ along with a persistent, nondifferentiable, and unknown ME.
After a coordinate transformation from a ME to a SD, it was shown that their stabilization method can be applied ensuring a decay of the CLF by introducing the bounds as $\bar e \leq \frac{\delta c}{2 \LL (2 + \Lf \delta)}$ and $\norm{d} \leq \frac{c}{2 \LL}$, where $c = \frac{\bar w}{2}$ (cf. decay in \cite{Kellett2004-weak}) is a bound of the decay and $d$ is the SD after the coordinate transformation.
This comes along with the imprecise strategy to "sample fast, but not too fast" \cite{Kellett2004-weak}.
However, if MEs and SDs appear simultaneously, the bounds are mixed up.
To demonstrate this, consider $\dot x = f(x, \kappa(x_k + E_k)) + q$ with $\hat x_k = x_k + E_k$.
Let $e$ be a globally Lipschitz continuous function with the property $e(k \delta) = E_k$ for all $k \geq 0$.
Apply the coordinate transformation $\xi := x + e$ (cf. \cite{Kellett2004-weak}), such that 
\begin{equation}
	\begin{split}
		\dot \xi &= f(\xi - e, \kappa(\xi)) + q + \dot e \\
		&= f(\xi, \kappa(\xi)) + \underbrace{f(\xi - e, \kappa(\xi)) - f(\xi, \kappa(\xi)) + q + \dot e}_{=:d}
	\end{split}
\end{equation}
holds.
The new SD $d$ can be bounded as 
\begin{equation}
	\norm{d} \leq \Lf \norm{e} + \norm{\dot e} + \norm{q} \leq \Lf \bar e + \frac{2 \bar e}{\delta} + \bar q,
\end{equation}
where $\frac{2 \bar e}{\delta}$ is a Lipschitz constant of $e$. 
To obtain the result in \cite{Kellett2004-weak}, $\norm{q}$ has to be bounded as $\norm{d} \leq \frac{\bar w}{4 \LL}$.
Ensuring 
\begin{equation}
	\begin{split}
		&\Lf \bar e + \frac{2 \bar e}{\delta} + \bar q \stackrel{!}{\leq} \frac{\bar w}{4 \LL} \\
		&\Leftrightarrow \bar e \leq \frac{\delta \bar w}{4 \LL (2 + \delta \Lf)} - \frac{\delta}{2 + \delta \Lf} \bar q.
	\end{split}
\end{equation}

satisfies $\Lf \bar e + \frac{2 \bar e}{\delta} + \bar q \leq \frac{\bar w}{4 \LL}$. 
Thus, both bounds $\bar e$ and $\bar q$ are connected and can not be chosen independently.

	
\subsection{Robustness of optimization-based control} \label{sec:rob-opt-bsd-ctrl}

Since OBC combines \eqref{eqn:Dini-aim-step-1} and \eqref{eqn:Dini-aim-step-2} in one step, the problems of DiA in stabilization under MEs and SDs do not appear.
The following lemma shows how the decay changes in this case.

\begin{lem} \label{lem:opt-bsd-ctrl-rob}
	Consider \eqref{eqn:sys-SH} along with a computed optimal control \eqref{eqn:opt-bsd-ctrl-law} and let Assumption \ref{asm:homog-asm} hold.
	The determined control law is robust w.r.t. SDs, i.e., $\dot x = f(x,u) + q$, as well as MEs, i.e., $\dot x = f(x + e,u)$, which means that \eqref{eqn:opt-bsd-ctrl-law} robustly practically stabilizes the origin of both systems.
\end{lem}
\begin{proof}
\textbf{Part 1:} Robustness w.r.t. SD: \\
	Using Lipschitz continuity of $L$, the decay is obtained as:
		\begin{equation} \label{eqn:OBC-robust-part-1}
			\begin{split}
				&L(x_k + \delta (f(x_k, \kappa(x_k)) + q)) - L(x_k) \\
				&\leq \LL \norm{\delta (f(x_k, \kappa(x_k)) + q)} \\
				&\leq \LL \delta \norm{f(x_k, \kappa(x_k))} + \LL \delta \norm q 
				\leq - \delta \frac{\bar w}{2} + \delta \LL \bar q.
			\end{split}
		\end{equation}
		The first term in the last line in \eqref{eqn:OBC-robust-part-1} represents the decay in the nominal case (cf. Theorem \ref{thm:stabilization-OBC}).
		The second one allows to introduce an upper bound $\bar q$ on the SD $q$.
		Choose $\bar q \leq \frac{\bar w}{4 \LL}$ to obtain $L(x_k + \delta (f(x_k, \kappa(x_k)) + q)) - L(x_k) \leq - \delta \frac{\bar w}{4}$ as a decay.
		Finally, Lemma \ref{lem:conv-target-ball} is applied to show the convergence of the closed-loop trajectories into the target ball $\ball_s$.
		
	\textbf{Part 2:} Robustness w.r.t. ME: \\
	In the case of a ME, the Lipschitz continuity of \eqref{eqn:sys} is used:
		\begin{equation} \label{eqn:OBC-robust-part-2}
			\begin{split}
				&L(x_k + e_k + \delta f(x_k + e_k, \kappa(\hat x_k))) - L(x_k) \\
				&\leq \LL \norm{e_k + \delta f(x_k + e_k, \kappa(\hat x_k))} \\
				&\leq \LL \norm{e_k} + \LL \delta \norm{f(x_k + e_k, \kappa(\hat x_k))} \\
				&\leq \LL \bar e + \LL \delta \norm{f(x_k, \kappa(\hat x_k)))} \\
				& \ \ \ + \delta \underbrace{\LL \norm{f(x_k + e_k, \kappa(\hat x_k))) - f(x_k, \kappa(\hat x_k)))}}_{\leq \LL \Lf \norm{e_k} \leq \LL \Lf \bar e} \\
				&\leq - \delta \frac{\bar w}{2} + \LL \left(1 + \delta \Lf \right) \bar e.
			\end{split}
		\end{equation}
		Again, the first term represents the decay of the nominal system and the second one allows to determine a bound for $\bar e$.
		The decay 
		\begin{equation}
			L(x_k + e_k + \delta f(x_k + e_k, \kappa(\hat x_k)))) - L(x_k) \leq - \delta \frac{\bar w}{4}
		\end{equation}
		is obtained by choosing $\bar e \leq \delta \frac{\bar w}{4 \LL \left(1 + \delta \Lf \right)}$.
		Lemma \ref{lem:conv-target-ball} ensures that the closed-loop trajectories converge into $\ball_s$.		
\end{proof}

\subsection{Robustness of infimum-convolution stabilization} \label{sec:rob-inf-con-stab}

The following lemma is given for the sake of completeness.
Further properties on InfC can be found in standard literature of nonsmooth analysis \cite{Clarke2004-lyapunov}, \cite{Clarke2011-discont-stabilization}, \cite{Clarke1997-stabilization}, \cite{Sontag1999-stability-disturb}, where robustness of InfC stabilization was also investigated.
\begin{lem} \label{lem:InfC-rob}
	Consider the system \eqref{eqn:sys} and let Assumption \ref{asm:sys-props} hold.
	Let $L$ be a CLF satisfying \eqref{eqn:decay}. 
	Then, the origin of \eqref{eqn:sys} can be practically robustly stabilized by an InfC control in the SH mode \eqref{eqn:sys-SH} in the sense of Definition \ref{def:robust-pract-stab}.
\end{lem}
The proof can be found in \cite[Theorem 1]{Schmidt2021-inf}, where Assumption \ref{asm:homog-asm} was only necessary for computational uncertainties.
\begin{rem}
	The proof of Lemma \ref{lem:InfC-rob} yields bounds for $\bar e$ and $\bar q$.
	The latter depends on the InfC parameter $\alpha$ (cf. \cite[Eqn. (43) and (45)]{Schmidt2021-inf}), which yields the same issues as the bound for $\bar e$ in DiA.
	Combining the four different bounds for $\bar q$ in Part 4 of the proof in \cite{Schmidt2021-inf} yields $\bar q \leq c_1 \bar w \alpha + c_2 \Lf$ with $c_1, c_2 > 0$, which depend on a bound of the system dynamics, $\chi$ (cf. \eqref{eqn:homog-asm-eq-2}) as well as some upper bound for the values of the CLF.
	In \cite[Eqn. (56)]{Schmidt2021-inf}, where the CLF is denoted as $V$, the term $L_\alpha(\hat x) - L_\alpha(x)$ is bounded from above by $\LL \delta \bar e$ and yields a bound for $\bar e$.
	Based on Assumption \ref{asm:sys-props}, $L_\alpha(\hat x) - L_\alpha(x) \leq \LL \bar e$ holds, such that $\bar e \leq \frac{1}{16} \delta \frac{\bar w}{\LL}$ is chosen.
	It includes $\delta$ meaning the robustness properties explicitly depend on the control parameters, like in DiA.
\end{rem}

An overview of the three stabilization methods and their robustness properties w.r.t. MEs and SDs is shown in Table \ref{tab:overview-stab-methods}.
Since $\bar e$ and $\bar q$ are connected in DiA, separate bounds can not be given.
Furthermore, the bounds for the SDs do not depend on $\delta$, whereas the bounds of the MEs do.
All these bounds are upper bounds for the ME and the SD to ensure a decay of the CLF ensuring a convergence into the target ball if they hold.
These bounds are difficult to compare directly, since in OBC, for example, the bounds for $\bar e$ and $\bar q$ can be derived in different ways.
To obtain the same decay of the CLF in OBC as in InfC \cite{Schmidt2021-inf}, namely $L(x_{k + 1}) - L(x_k) \leq - \frac 3 8 \delta \bar w$ outside the target ball $\ball_s$, the bounds have to be chosen to satisfy $- \delta \frac{\bar w}{2} + \LL (1 + \delta \Lf) \bar e + \delta \LL \bar q \leq - \frac 3 8 \delta \bar w$ (cf. \eqref{eqn:OBC-robust-part-1} and \eqref{eqn:OBC-robust-part-2}).
Assuming the same bound for $\bar e$ as in InfC (see Table \ref{tab:overview-stab-methods}), $\bar q \leq \frac{\bar w}{\delta \LL} \left( \frac \delta 8 - \frac{1 + \delta \Lf}{16} \right)$ in OBC is required.

The performance of the three different methods is compared in the following case study.

\begin{table}
		\caption{Overview on robustness properties of the nonsmooth stabilization methods for the system $\dot x = f(x, \kappa(x + e)) + q$.}
		\label{tab:overview-stab-methods}
		\renewcommand{\arraystretch}{1.5}
		\begin{tabularx}{0.49\textwidth}{|l||X|X|}
		\hline
		Method & Measurement error (ME) $\bar e$ & System disturbance (SD) $\bar q$ \\ \hline \hline
		DiA & \multicolumn{2}{c|}{$\bar e \leq \frac{\delta \bar w}{4 \LL (2 + \delta \Lf)} - \frac{\delta}{2 + \delta \Lf} \bar q$} \\ \hline
		OBC & $\bar e \leq \delta \frac{\bar w}{4 \LL \left(1 + \delta \Lf \right)}$ & $\bar q \leq \frac{\bar w}{4 \LL}$ \\ \hline
		InfC & $\bar e \leq \delta \frac{\bar w}{16 \LL}$ & $\bar q \leq c_1 \bar w \alpha + c_2 \Lf$ \\ \hline
		\end{tabularx}				
\end{table}

\section{Case study} \label{sec:case-study}

%

A well-known example for a system that fails to satisfy Brockett's conditions \cite{Brockett1983-stabilization} is the nonholonomic integrator
\begin{equation}
	\label{eqn:NI}
	\tag{NI}
	\dot \varphi = \begin{pmatrix}
		1 \\ 0 \\ -\varphi_2
	\end{pmatrix} \omega_1 + \begin{pmatrix}
		0 \\ 1 \\ \varphi_1
	\end{pmatrix} \omega_2 = g_1(\varphi) \omega_1 + g_2(\varphi) \omega_2,
\end{equation}
also known as Brockett's integrator.
Adding additional integrators right before the control inputs yields the extended nonholonomic double integrator (ENDI) \cite{Schmidt2021-inf}
\begin{equation} 
	\label{eqn:ENDI}
	\tag{ENDI}
	\begin{aligned}
		\dot \varphi_1 &= \eta_1, & \dot \varphi_2 &= \eta_2, & \dot \varphi_3 &= \varphi_1 \eta_2 - \eta_1 \varphi_2, \\
		\dot \eta_1 &= u_1, & \dot \eta_2 &= u_2. \\
	\end{aligned}
\end{equation}
In \cite{Osinenko2020-nonsmooth}, nonsmooth backstepping was applied to \eqref{eqn:ENDI} based on the idea of \cite{Matsumoto2015-position}, where a semiconcave CLF of the NI is written as the minimum w.r.t. a parameter $\lambda\in \Lambda$, $\Lambda \subseteq \R^{2n}$ of a function $F$.
This function $F$ is continuously differentiable for each $\lambda \in \Lambda$ \cite{Cannarsa2004-semiconcave} and is used in the traditional backstepping technique to obtain the CLF with state vector $x = \begin{pmatrix} \varphi^\top & \eta^\top \end{pmatrix}^\top$ as
\begin{equation} 
	\label{eqn:ENDI-V}
	L(x) = \min_{\lambda \in [0,2 \pi)} \left\{  \tilde F(\varphi; \lambda) + \frac{1}{2} \norm{ \eta - \kappa(\varphi; \lambda) }^2 \right\},
\end{equation}
where 
\begin{equation}
	\tilde F(\varphi; \lambda) = \varphi_1^2 + \varphi_2^2 + 2 \varphi_3^2 - 2 \varphi_3 (\varphi_1 \cos \lambda + \varphi_2 \sin \lambda)
\end{equation}
and
\begin{equation}
	\kappa(\varphi; \lambda) = -\begin{pmatrix} \scal{\nabla_\varphi \tilde F(\varphi; \lambda), g_1(\varphi)} \\ \scal{\nabla_\varphi \tilde F(\varphi; \lambda), g_2(\varphi)} \end{pmatrix}.
\end{equation}
Using the minimizer $\lambda^\star$ of \eqref{eqn:ENDI-V}, the CLF of the NI given in \cite{Braun2017-SH-stabilization-Dini-aim} is obtained: 
\begin{equation}
	\tilde F(\varphi; \lambda^\star) = \tilde L(\varphi) = \varphi_1^2 + \varphi_2^2 + 2 \varphi_3^2 - 2 \abs{\varphi_3} \sqrt{ \varphi_1^2 + \varphi_2^2 }.
\end{equation}
The obtained CLF fulfills Assumption \ref{asm:homog-asm}, which can be shown by a short computation (cf. \cite[Example 1]{Osinenko2020-nonsmooth}).

In the following, the influences of $\delta$, $\bar e$, and $\bar q$ are investigated.
Upper bounds on $\delta$ were computed in the case study in \cite{Schmidt2021-inf} resulting in a conservative choice such that no significant change in the trajectories could be seen in this case.
Therefore, the values $\delta \in \{ 0.25, 0.5 \}$ are selected along with $\bar e, \bar q \in \{ 0.01, 0.1 \}$.
The closed-loop trajectories $\hat x^{\text{DiA}}$, $\hat x^{\text{OBC}}$, and $\hat x^{\text{InfC}}$ are compared, which are obtained by applying the determined control to \eqref{eqn:ENDI} in the SH manner.
Twenty initial values with the same norm $S$ are generated for each method to exclude biased results.
To compare the results with those in \cite{Schmidt2021-inf}, we set $S := \norm{\begin{pmatrix} -1 & 0.5 & 0.2 & 0.1 & 0.1 \end{pmatrix}}_2$ and choose the input constraints as $\U = [-1,1]^2$.
The parameters of DiA and InfC are chosen as $r = 0.1$ and $\alpha = 0.1$, respectively.

\begin{figure}
	\centering
		\includegraphics[width = 0.48\textwidth, trim={0.2cm 0.25cm 0.2cm 0.25cm},clip]{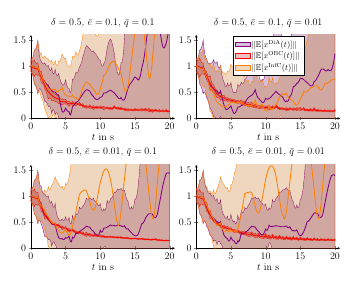}
		\caption{Norm of the mean and the standard deviation of the trajectories $x^i(t)$ for different MEs $\bar e$ and SDs $\bar q$ for sampling period $\delta = 0.5$.}
		\label{fig:result-example-delta1-graphs}
\end{figure}

\begin{figure}				
		\includegraphics[width = 0.48\textwidth, trim={0.2cm 0.25cm 0.2cm 0.25cm},clip]{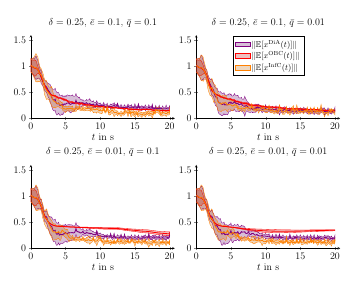}		
		\caption{Norm of the mean and the standard deviation of the trajectories $x^i(t)$ for different MEs $\bar e$ and SDs $\bar q$ for sampling period $\delta = 0.25$.}
		\label{fig:result-example-delta2-graphs}
\end{figure}

Fig. \ref{fig:result-example-delta1-graphs} and Fig. \ref{fig:result-example-delta2-graphs} shows the norm of the mean (solid lines) and the standard deviation (shaded areas) of closed-loop trajectories for $\delta = 0.5$ and $\delta = 0.25$.
OBC is clearly the best method to stabilize \eqref{eqn:ENDI} in the presence of large MEs, SDs, and sampling periods.
For DiA and InfC, the closed-loop trajectories oscillate with high amplitudes failing to stabilize \eqref{eqn:ENDI} at the origin.
However, DiA performs better than InfC.
Increasing the measurement frequency to $\delta = 0.25$ results in a better performance of the each method, where DiA and especially InfC are able to provide even better results than OBC for small MEs and SDs s.t. the closed-loop trajectories remain in a closer surrounding of the origin. 
In conclusion, OBC along with its robustness w.r.t. MEs and SDs provides the best results, since the computations are also the fastest based on empirical investigations, presumable due to the usage of one step in computing the control action in comparison to DiA and InfC.
Table \ref{tab:computation-times} shows the average times to compute the stabilizing control for a single initial value.
\begin{table}
		\renewcommand{\arraystretch}{1.1}
		\caption{Comparison of computation times of the methods.}
		\begin{tabularx}{0.49\textwidth}{|l||X|X|X|}
		\hline
		Method & DiA & OBC & InfC \\ \hline \hline
		$\delta = 0.5$ & 83.1 s & 28.4 s & 80.8 s \\ \hline
		$\delta = 0.25$ & 95.7 s & 39.9 s & 105.1 s \\ \hline
		\end{tabularx}			
		\label{tab:computation-times}		
\end{table}

Tables \ref{tab:result-example-delta2} and \ref{tab:result-example-delta3} show the mean and the standard deviation for measurement error bounds $\bar e \in \{ 0, 0.01, 0.1, 1 \}$ and system disturbance bounds $\bar q \in \{ 0, 0.01, 0.1, 1 \}$ for sampling times $\delta = 0.5$ and $\delta = 0.25$.
With the exception of a few anomalies, it can be seen that the smaller $\bar e$ and $\bar q$ are, the smaller are mean and standard deviation of the different trajectories.
The same holds for $\delta$, since the lower the sampling time is, the higher are mean and standard deviation. 
A closer look on the performance of the three stabilization methods shows that DiA works better than InfC for high measurement errors, which can be seen by comparing the first lines in the respective tables.
This shows the mentioned abnormal robustness w.r.t. measurement errors (cf. Section \ref{sec:rob-Dini-aiming}).
For smaller $\bar e$, however, InfC is better than DiA, but worse than OBC, which has the best performance of these three methods.
It can be seen that mean and standard deviation in OBC is partially significantly smaller in comparison with the two other methods - especially if the measurement error or the system disturbance is large.
This robustness of OBC w.r.t. measurement errors and system disturbance was already shown in Section \ref{sec:rob-opt-bsd-ctrl}.

\begin{table}
		\caption{Comparison of different bounds for measurement error $\bar e$ and system disturbance $\bar q$ for sampling period $\delta = 0.5$ with 20 samples w.r.t. mean (M) and standard deviation (D).}
		\begin{tabularx}{0.49\textwidth}{|X||X|X|X|X|}
		\hline
			DiA, M & $\bar q = 1$ & $\bar q = 0.1$ & $\bar q = 0.01$ & $\bar q = 0$ \\ \hline \hline
			$\bar e = 1$ & 7.8382 & 6.1745 & 6.8838 & 5.8479 \\ \hline
    		$\bar e = 0.1$ & 1.4944 & 0.5827 & 0.5114 & 0.8455 \\ \hline
			$\bar e = 0.01$ & 0.6911 & 0.4205 & 0.4106 & 0.4111\\ \hline
			$\bar e = 0$ & 0.4106 & 0.4106 & 0.4106 & 0.4106 \\ \hline
		\end{tabularx}
		\vspace{0.1cm} \\
		\begin{tabularx}{0.49\textwidth}{|X||X|X|X|X|}
		\hline
			DiA, D & $\bar q = 1$ & $\bar q = 0.1$ & $\bar q = 0.01$ & $\bar q = 0$ \\ \hline \hline
			$\bar e = 1$ & 2.2515 & 1.4414 & 2.0497 & 1.6949 \\ \hline
			$\bar e = 0.1$ & 0.5218 & 0.1723 & 0.1873 & 0.2348 \\ \hline
			$\bar e = 0.01$ & 0.2919 & 0.1967 & 0.2012 & 0.2015 \\ \hline
			$\bar e = 0$ & 0.2012 & 0.2012 & 0.2012 & 0.2012 \\ \hline
		\end{tabularx}		
		\vspace{0.1cm} \\
		\begin{tabularx}{0.49\textwidth}{|X||X|X|X|X|}
		\hline
			OBC, M & $\bar q = 1$ & $\bar q = 0.1$ & $\bar q = 0.01$ & $\bar q = 0$ \\ \hline \hline
			$\bar e = 1$ & 1.0067 & 1.2809 & 1.1684 & 0.7606 \\ \hline
			$\bar e = 0.1$ & 0.2992 & 0.2176 & 0.2225 & 0.2231 \\ \hline
			$\bar e = 0.01$ & 0.3585 & 0.1904 & 0.2006 & 0.2126 \\ \hline
			$\bar e = 0$ & 0.2922 & 0.1939 & 0.2184 & 0.2210 \\ \hline
		\end{tabularx}
		\vspace{0.1cm} \\
		\begin{tabularx}{0.49\textwidth}{|X||X|X|X|X|}
		\hline
			OBC, D & $\bar q = 1$ & $\bar q = 0.1$ & $\bar q = 0.01$ & $\bar q = 0$ \\ \hline \hline
			$\bar e = 1$ & 0.2299 & 0.2959 & 0.2989 & 0.1881 \\ \hline
			$\bar e = 0.1$ & 0.1001 & 0.0338 & 0.0367 & 0.0352 \\ \hline
			$\bar e = 0.01$ & 0.1364 & 0.0239 & 0.0186 & 0.0214 \\ \hline
			$\bar e = 0$ & 0.0859 & 0.0225 & 0.0214 & 0.0262 \\ \hline
		\end{tabularx}		
		\vspace{0.1cm} \\
		\begin{tabularx}{0.49\textwidth}{|X||X|X|X|X|}
		\hline
			InfC, M & $\bar q = 1$ & $\bar q = 0.1$ & $\bar q = 0.01$ & $\bar q = 0$ \\ \hline \hline
			$\bar e = 1$ & 5.7924 & 4.8118 & 4.6637 & 4.8519 \\ \hline
			$\bar e = 0.1$ & 1.8177 & 1.5791 & 1.3521 & 1.1839 \\ \hline
			$\bar e = 0.01$ & 0.7716 & 0.7755 & 0.7872 & 0.7320 \\ \hline
			$\bar e = 0$ & 0.7084 & 0.7084 & 0.7084 & 0.7084 \\ \hline
		\end{tabularx}
		\vspace{0.1cm} \\
		\begin{tabularx}{0.49\textwidth}{|X||X|X|X|X|}
		\hline
			InfC, D & $\bar q = 1$ & $\bar q = 0.1$ & $\bar q = 0.01$ & $\bar q = 0$ \\ \hline \hline
			$\bar e = 1$ & 1.7492 & 1.3098 & 1.2638 & 1.2542 \\ \hline
			$\bar e = 0.1$ & 0.5051 & 0.4596 & 0.3352 & 0.3004 \\ \hline
			$\bar e = 0.01$ & 0.1750 & 0.1764 & 0.1772 & 0.1747 \\ \hline
			$\bar e = 0$ & 0.1731 & 0.1731 & 0.1731 & 0.1731 \\ \hline
		\end{tabularx}					
		\label{tab:result-example-delta2}
\end{table}

\begin{table}
		\caption{Comparison of different bounds for measurement error $\bar e$ and system disturbance $\bar q$ for sampling period $\delta = 0.25$ with 20 samples w.r.t. mean (M) and standard deviation (D).}
		\begin{tabularx}{0.49\textwidth}{|X||X|X|X|X|}
		\hline
			DiA, M & $\bar q = 1$ & $\bar q = 0.1$ & $\bar q = 0.01$ & $\bar q = 0$ \\ \hline \hline
			$\bar e = 1$ & 0.5423 & 0.5206 & 0.4392 & 0.4477 \\ \hline
			$\bar e = 0.1$ & 0.3019 & 0.2980 & 0.3295 & 0.3259  \\ \hline
			$\bar e = 0.01$ & 0.3825 & 0.3841 & 0.3833 & 0.3749 \\ \hline
			$\bar e = 0$ & 0.3847 & 0.3847 & 0.3847 & 0.3847 \\ \hline
		\end{tabularx}
		\vspace{0.1cm} \\
		\begin{tabularx}{0.49\textwidth}{|X||X|X|X|X|}
		\hline
			DiA, D & $\bar q = 1$ & $\bar q = 0.1$ & $\bar q = 0.01$ & $\bar q = 0$ \\ \hline \hline
			$\bar e = 1$ & 0.1401 & 0.1481 & 0.1219 & 0.1247 \\ \hline
			$\bar e = 0.1$ & 0.0539 & 0.0507 & 0.0490 & 0.0556 \\ \hline 
			$\bar e = 0.01$ & 0.0446 & 0.0447 & 0.0446 & 0.0469  \\ \hline
			$\bar e = 0$ & 0.0436 & 0.0436 & 0.0436 & 0.0436 \\ \hline
		\end{tabularx}		
		\vspace{0.1cm} \\
		\begin{tabularx}{0.49\textwidth}{|X||X|X|X|X|}
		\hline
			OBC, M & $\bar q = 1$ & $\bar q = 0.1$ & $\bar q = 0.01$ & $\bar q = 0$ \\ \hline \hline
			$\bar e = 1$ & 0.5290 & 0.5180 & 0.4114 & 0.4187\\ \hline
			$\bar e = 0.1$ & 0.2383 & 0.1887 & 0.1872 & 0.1872  \\ \hline
			$\bar e = 0.01$ & 0.2845 & 0.1934 & 0.1861 & 0.1973  \\ \hline
			$\bar e = 0$ & 0.2442 & 0.1684 & 0.1917 & 0.1963 \\ \hline
		\end{tabularx}
		\vspace{0.1cm} \\
		\begin{tabularx}{0.49\textwidth}{|X||X|X|X|X|}
		\hline
			OBC, D & $\bar q = 1$ & $\bar q = 0.1$ & $\bar q = 0.01$ & $\bar q = 0$ \\ \hline \hline
			$\bar e = 1$ & 0.1333 & 0.1567 & 0.1144 & 0.1033 \\ \hline
			$\bar e = 0.1$ & 0.0480 & 0.0256 & 0.0238 & 0.02171 \\ \hline
			$\bar e = 0.01$ & 0.0607 & 0.0071 & 0.0028 & 0.0010\\ \hline
			$\bar e = 0$ & 0.0620 & 0.0071 & 0.0033 & 1.8 $\cdot 10^{-6}$ \\ \hline
		\end{tabularx}		
		\vspace{0.1cm} \\
		\begin{tabularx}{0.49\textwidth}{|X||X|X|X|X|}
		\hline
			InfC, M & $\bar q = 1$ & $\bar q = 0.1$ & $\bar q = 0.01$ & $\bar q = 0$ \\ \hline \hline
			$\bar e = 1$ & 1.4121 & 1.2769 & 0.7809 & 0.8154 \\ \hline 
			$\bar e = 0.1$ & 0.2790 & 0.2657 &  0.2613 & 0.2747 \\ \hline
			$\bar e = 0.01$ & 0.2996 & 0.2536 & 0.2695 & 0.2359 \\ \hline
			$\bar e = 0$ & 0.2581 & 0.2581 & 0.2581 & 0.2581 \\ \hline
		\end{tabularx}
		\vspace{0.1cm} \\
		\begin{tabularx}{0.49\textwidth}{|X||X|X|X|X|}
		\hline
			InfC, D & $\bar q = 1$ & $\bar q = 0.1$ & $\bar q = 0.01$ & $\bar q = 0$ \\ \hline \hline
			$\bar e = 1$ & 0.3441 & 0.3250 & 0.1602 & 0.1678 \\ \hline 
			$\bar e = 0.1$ & 0.0891 & 0.0787 & 0.0825 & 0.0851 \\ \hline
			$\bar e = 0.01$ & 0.0613 & 0.0911 & 0.0667 & 0.0626 \\ \hline
			$\bar e = 0$ & 0.0812 & 0.0812 & 0.0812 & 0.0812 \\ \hline
		\end{tabularx}					
		\label{tab:result-example-delta3}
\end{table}

\section{Conclusion} \label{sec:concl}

This publication investigated three different nonsmooth stabilization techniques, namely Dini aiming, optimization-based control, and stabilization via infimum convolutions.
Selected robustness properties of these approaches were extensively studied formally and in simulation.

\bibliographystyle{plain}        
\bibliography{bib/constructing-LFs,bib/discont-DE,bib/stabilization,bib/stability,bib/non-smooth-analysis,bib/sliding-mode,bib/MPC,bib/opt-ctrl,bib/dyn-sys,bib/semiconcave}

%
\end{document}